\documentclass[a4paper,11pt]{article}

\usepackage{jheppub} 
\usepackage{graphicx, graphics, color}
\usepackage[T1]{fontenc} 

\usepackage{amsmath, amssymb, bm, graphicx, graphics, color, mathrsfs}

\newcommand{\be}{\begin{equation}}
\newcommand{\ee}{\end{equation}}
\newcommand{\ben}{\begin{eqnarray}}
\newcommand{\een}{\end{eqnarray}}

\newcommand{\la}{{\lambda}}

\newcommand{\cO}{{\cal O}}

\newcommand{\p}{\partial}
\newcommand{\na}{\nabla}

\newcommand{\ep}{\epsilon}

\newcommand{\ga}{\gamma}

\title{\boldmath 
Viscosity of holographic fluid in the presence of dark matter sector}

\author[1]{Marek Rogatko\note{rogat@kft.umcs.lublin.pl, marek.rogatko@poczta.umcs.lublin.pl}}
\author[2]{Karol I. Wysokinski\note{karol@tytan.umcs.lublin.pl}}
\affiliation{Institute of Physics \\
Maria Curie-Sk{\l}odowska University \\
20-031 Lublin, pl. Marii Curie-Sk{\l}odowskiej 1, Poland}

\abstract{
Based on the gauge/gravity correspondence, the  hydrodynamic response coefficients, 
shear and Hall viscosities, have been studied.
The holographic model of Einstein-Maxwell- AdS $(3+1)$-dimensional system additionally 
coupled with the another gauge field mimicking the {\it dark matter} sector, 
as well as, gravitational Chern-Simons term bounded with a dynamical scalar 
field, were taken into account.
Condensation of the scalar field in the presence of the 
 deformation chemical potential for the {\it dark matter} gauge field
provide the  parity violating terms. Both shear and Hall viscosities have been 
calculated and their dependence on $\alpha$ - the coupling constant between matter 
and {\it dark matter} sectors has been studied. To the lowest
order in the derivative expansion and  perturbation in $\alpha$, the shear viscosity
is not influenced by the {\it dark matter}, while the Hall component linearly depends on $\alpha$.\\ } 


\begin{document} 

\maketitle
\flushbottom


\section{Introduction}
\label{sec:intro}
The exploit of the gauge/gravity correspondence \cite{mal,wit98,gub98} in studying strongly correlated systems 
resulted, among others, in establishing the lower bound $\hbar/4\pi$ on the ratio of the shear 
viscosity $\eta_s$ to entropy density $s$ in holographic fluid \cite{kovtun2005}. This interesting
result  has contributed  to the  deeper understanding of the state of strongly interacting 
quark-gluon plasma obtained at RHIC \cite{rom07,mat07}. 
Related studies based on the gauge/gravity duality \cite{hartnoll2007} have also triggered the shear viscosity 
measurements in the  ultracold Fermi gases \cite{cao11}, and more recently in the condensed matter systems such as  
graphene~\cite{crossno2016,bandurin2016} and strongly correlated  oxide \cite{moll2016}. 
The comprehensive discussion of this novel set of experiments is given in~\cite{zaanen2016}.

The desire to understand the long wave-length particle behavior {\it via} hydrodynamic analogy has a long history.
It goes back to the Madelung's hydrodynamic formulation of quantum mechanics~\cite{madelung1926}, which later has been
applied to many different condensed matter systems including quantum wires~\cite{dej1995}, two-dimensional
weakly~\cite{eaves1999} and strongly~\cite{abanov2013} interacting electron gas in magnetic 
field and even nanoscale conductors~\cite{dagosta2006}. 
The early proposal in the field of elementary particle physics~\cite{landau1953,bel56} due to Landau, 
suggesting such description of hadronic fireballs also has to be mentioned in this context. 

Recently, there has been a great resurgence of interest in the description of viscosity 
components in fluids by means of quantum theory \cite{avr95,rea11,bra12} or gauge/gravity 
duality \cite{jan06a,jan06b,jan07,sar06,buc05,ben06,buc08,buc08a,buc08b,buc08c,mye09,cre11,bri08,bri08a,reb12,jai15,ge15,
alb16,sar12,che12,roy14,cai12,zou14,liu13,liu14,liu14a,son14}. The problem was elaborated from 
various points of view including the universality of the conjectured lower bound on the ratio $\eta_s/s$ 
and its measurements in an interacting gas of fermions  near the unitarity limit \cite{cao11}. Various hydrodynamic
response functions have been probed using AdS/CFT approach~\cite{erd13}-\cite{jen12a}. Among them the antisymmetric part of the
viscosity tensor, the so-called Hall viscosity analogous  to Hall conductivity has attracted a lot of attention.
The Hall viscosity being non-dissipative viscosity coefficient does not contribute to the entropy 
production of the fluid. In quantum fluids, at zero temperature, the dissipative shear and bulk viscosities disappear.
On the contrary, non-dissipative Hall viscosity can remain nonzero  in systems with broken parity or 
 time-reversal symmetry~\cite{sar12}. There is, however a problem with this component of the 
viscosity related to the the question of how to measure it in condensed matter systems~\cite{hua15}. 

The gauge/gravity correspondence offers quite a deep insight into the problem of understanding 
 the dynamics of strongly interacting systems. 
However, one has to remember that in the hydrodynamic description the universal 
ratio~\cite{sar06,buc05,ben06,buc08,buc08a,buc08b,buc08c,mye09,cre11}, 
 strictly speaking, is valid for all gauge
theories with Einstein gravity in the limit $N \rightarrow \infty$ and $\la \rightarrow \infty$, 
where $N$ is the numbers of colors, while $\la$ stands for t'Hooft coupling. 
The division of the $\eta_s$ by entropy density allows to get rid of the number 
of degrees of freedom and obtain the universal bound.
On the other hand, it turns out that in higher derivatives theories the aforementioned bound is 
not universal~\cite{bri08}. The anisotropic theories also allow 
the bound violation \cite{reb12,jai15,ge15}. The violation of the  viscosity bound has also been predicted
in massive~\cite{alb16,sad16} and in the quadratic Gauss-Bonnet gravity~\cite{bri08a}.  In the later work it has been 
noted that the field excitations in the dual field theory enable the superluminal propagation 
velocities for the Gauss-Bonnet coupling constant greater than $9/100$.

The techniques developed in the AdS/CFT correspondence enable studies of parity violating 
effects in hydrodynamical systems at strong coupling \cite{sar12,che12,roy14,cai12,zou14,liu13,liu14,liu14a,son14}. 
In general the systems with parity violations acquire additional response parameter/transport coefficient 
in the long wave-length limit. As it is well known from classic physics the asymmetric  component of the
viscosity appears  at the same order as the shear viscosity in the hydrodynamical derivative expansion. 
It is subject to the parity or time reversal violation. In condensed matter physics 
it is called Hall viscosity~\cite{avr95} and we adopt this name in the following. 

The holographic model of Hall viscosity in $(2+1)$-dimensional system was given in \cite{sar12}, 
where the dynamical Chern-Simons term was used in the calculations.
The further generalizations, both analytical and numerical, in the model with Chern-Simons 
and Maxwell terms \cite{che12}, as well as, the Born-Infeld black branes \cite{roy14}
were presented. The Hall  and shear viscosities, in the model with 
dynamical Chern-Simons terms, were elaborated in \cite{cai12,zou14}.
The spontaneously generated angular momentum in models with gauge and Chern-Simons 
terms were studied in \cite{liu13,liu14a}. Moreover, 
recently it was reported that the ratio between Hall viscosity and angular momentum 
density is a constant, at least near the critical regime \cite{son14}. It was envisaged in the holographic
$p_x +ip_y$ model, which is different from the Chern-Simons one studied before. 
The ground state breaks spatial parity by locking it to non-Abelian gauge parity,  
which in turn is broken by the appearance of non-Abelian gauge connection, the only source of the Hall viscosity,
angular momentum density and Hall conductivity.

The studies of hydrodynamic response via gauge/gravity duality is of interest {\it per se}. The 
additional motivation behind our work here is related to the widely debated issue of the 
invisible component of the  matter in the Universe, the so called {\it dark matter} and
its experimental detection. In the previous studies~\cite{nak14,nak15,nak15a1,rog15,rog15a} we have analyzed the
properties of holographic superconductors and vortices with the hope to find such modifications of their properties 
which will allow the detection of the dark matter. The effect of {\it dark matter} on the properties 
of superconductors has also been discussed  in \cite{pen15a,pen15b}.

Here we extend the previous work to the analysis of a fluid response. 
To this end we  adopt the simple holographic realization of $(2+1)$-dimensional isotropic 
holographic fluid with spontaneously broken parity and additionally supplemented 
by the {\it dark matter} sector. The {\it dark matter} sector will be represented by the  
$U(1)$-gauge field which is coupled to the ordinary Maxwell one.

As was mentioned, the motivation standing behind our studies is to elucidate the imprint 
of the {\it dark matter}  on physical phenomena,
which detailed analysis would in turn allow to detect dark matter and thus answer  one of the most 
tantalizing questions of the contemporary physics. 
The present paper is a continuation and extension of the efforts aiming at 
elucidation of the effect of dark matter on the properties of condensed 
systems~\cite{nak14,nak15,nak15a1,rog15,rog15a,pen15a,pen15b}. 
The existing literature on the subject reports numerous theoretical and observational evidences
supporting the existence of dark matter and its role as a possible source of the observed 
anomalies~\cite{reg15,ali15,bra14,ful15,lop14,nak12,nak15a,ger15,til15,integral,atic,pamela,muon,massey15a,massey15b}.
Here we are interested in its influence on the viscosity of holographic fluids. 
In view of the laboratory experiments \cite{cao11,crossno2016,bandurin2016,moll2016} which outcomes seem to agree with the predictions of theories 
based on gauge/gravity duality we hope that future obsevations will find imprints of {\it dark matter} 
on the properties of quantum fluids. 

The paper is organized as follows. In section 2 we describe the main features 
of the model under inspection with two $U(1)$-gauge fields coupled together.  As already mentioned one of them is the ordinary
Maxwell field and the other is responsible for the  {\it dark matter} sector. General setup and the equations 
of motion are derived in section 3, while their solution to lowest order are presented in section 3.1. 
The effect of dark matter sector on shear and Hall viscosities is analyzed in section 4, while the subsection 4.1
is devoted to the analysis of the  temperature dependence of the Hall viscosity. We end up the paper with 
summary and conclusions (section 5).

\section{Holographic model}
\label{sec:model}
The gravitational background for the holographic model of viscosities constitutes 
the four-dimensional deformation of the general relativity, the so-called
Chern-Simons gravity, in the formulation proposed in \cite{jac03}.
Chern-Simons modified gravity authorizes the effective extension of Einstein 
theory taking into account gravitational parity violation. The aforementioned
extension is motivated by anomaly cancelation in string theory and particle physics. There were proposed 
some astrophysical tests of Chern-Simons modified gravity including
Solar system, binary pulsars, galactic rotation curves and gravitational wave experiments, as well as, 
the possible explanation of cosmological matter-anti matter asymmetry (for the 
contemporary review of the various aspects of the theory see~\cite{ale09}). Moreover, it turns out that the static solution of equation of motion in dynamical Chern-Simons gravity
with $U(1)$-gauge field is diffeomorphic to an open set of Reissner-Nordstr\"om non-extremal solution 
with electric charge \cite{rog13}.

The gravitational action in (3+1) dimensions is taken in the form
\be
S_{g} = \int \sqrt{-g}~ d^4 x~  \bigg( R - 2\frac{\Lambda}{L^2} - \frac{1}{2} \na_\mu \theta \na^\mu \theta - V(\theta) - \frac{\la}{4}~\theta~{}^{\ast}R~R
\bigg),
\label{sgrav} 
\ee
where $\theta$ is the pseudo scalar field, $\Lambda = - 3$ stands for the cosmological constant, L the radius of the
AdS space-time, which from now on  is taken as L=1. 
The Pontryagin density term and dual Riemann tensor are provided by
\be
{}^{\ast}R ~R = {}^{\ast}R^{\alpha}{}_{\beta}{}{}^{\gamma \delta}~R^{\beta}{}_{\alpha \gamma \delta}, \qquad
{}^{\ast}R^{\alpha}{}_{\beta}{}{}^{\gamma \delta} = \frac{1}{2} \ep^{\gamma \delta \eta \psi}~R^{\alpha}{}_{\beta \eta \psi}.
\ee
$\la$ is a coupling constant. 
The field $\theta$, sometimes called Chern-Simons coupling field, being a function of spacetime coordinates, 
serves as a deformation function. One can observe that when $\theta = 0$ and $V(\theta) =0$,
the above modification of gravity reduces to the Einstein theory. 
As was mentioned we treat $\theta$ as a pseudo scalar \cite{che12}, so the gravitational Chern-Simons term does not break the parity. 
But the pseudo scalar term violates parity spontaneously, as well as, enables to receive a pseudo scalar condensate at the boundary.


The potential $V(\theta)$ assumes standard Ginzburg-Landau form of the Mexican hat type for $m^2<0$ and $c>0$,
\be
V(\theta)= \frac{1}{2}~m^2 ~\theta^2 + \frac{1}{4}~c~\theta^4.
\ee
As was found in \cite{har08,fau09}  scalar field can develop an instability if its mass square violates the near horizon AdS Breitenlohner-Freedman  bound.
When the adequate condition is received, the scalar field will be stable at infinity but will condense near the event horizon of black brane.
It happens that the condensed solution will comprise nontrivial radial profile for scalar field. Namely,
at low temperatures the field in question, explores extreme values of its potential. Therefore the non-linearities in the potential connected with scalar field will be of a great importance.

At zero temperature, one expects that the $\theta$ field condenses until the value of it 
at the black brane event horizon reaches some point near
the bottom of the Mexican hat potential. At the aforementioned point the effective 
AdS-mass will fulfill the AdS Breitenlohner-Freedman  bound and the condensation will stop.
The condensation is expected to persist up to some finite temperature $T_c$, below which $\theta$ condensates. $m^2$ is the effective mass near the point $\theta=0$.
It ought to satisfy the Breitenlohner-Freedman condition.


The matter field is composed of the Abelian-Higgs sector coupled to the 
second $U(1)$-gauge field which in our theory describes the {\it dark matter} sector~\cite{bri11}.  
The  action incorporating {\it dark matter} is provided by 
\be
\label{s_matter}
S_{m} = \int \sqrt{-g}~ d^4x  \bigg( 
- \frac{1}{4}F_{\mu \nu} F^{\mu \nu} - \frac{1}{4} B_{\mu \nu} B^{\mu \nu} - \frac{\alpha}{4} F_{\mu \nu} B^{\mu \nu}
\bigg), 
\ee  
where
$F_{\mu \nu} = 2 \nabla_{[ \mu} A_{\nu ]}$ stands for the ordinary Maxwell field strength tensor, while
the second $U(1)$-gauge field $B_{\mu \nu}$ is given by $B_{\mu \nu} = 2 \nabla_{[ \mu} B_{\nu ]}$. 

It can be observed that the bulk action $S_g + S_m$ conserves parity, so a pseudo scalar $\theta$, in the last term of the equation (\ref{sgrav}), is of a key importance to introduce 
the parity violation in the boundary theory via $\theta$-condensation.

Let us comment on the motivation for introduction a {\it dark matter} sector in the form of equation (\ref{s_matter}). In our previous works dedicated to the subject of the {\it dark matter} influence on
holographic s-wave and p-wave superconductors we have established that the $\alpha$-coupling constant of Maxwell and $U(1)$-gauge {\it dark matter} fields influence
the various characteristics of the superconductors. If one treats the AdS/CFT correspondence as a kind of method enabling us insight into the properties of strongly correlated systems,
these changes may be treated as the guideline in future experiments detecting {\it dark matter}.

Of course, the action (\ref{s_matter}) can be rewritten in the form
\be
S_m = \int d^4x~\sqrt{-g} \bigg(
- \frac{1}{4} {\tilde F}_{\mu \nu}{\tilde F}^{\mu \nu} - \frac{1}{4} B'_{\mu \nu} B^{' \mu \nu} \bigg),
\ee
where we defined
\ben
A_{\mu} &=& \tilde A_\mu - \frac{\alpha}{2} B_\mu,\\
B'_{\mu \nu} & =& \sqrt{1- \frac{\alpha^2}{4}}~B_{\mu \nu},\\
{\tilde F}_{\mu \nu} &=& \na_{[ \mu }{\tilde A}_{\nu ]}.
\een
The other form of the action in question can be obtained when the ordinary Maxwell field is multiplied by a coefficient. It implies
\be
S_m = \int d^4x~\sqrt{-g} \bigg(
- \frac{1}{4} {F'}_{\mu \nu}{F'}^{\mu \nu} - \frac{1}{4} {\tilde B}_{\mu \nu} {\tilde B}^{ \mu \nu} \bigg),
\ee
where one defines the following:
\ben
B_{\mu} &=& {\tilde B}_\mu - \frac{\alpha}{2} A_\mu,\\
F'_{\mu \nu} & =& \sqrt{1- \frac{\alpha^2}{4}}~F_{\mu \nu},\\
{\tilde B}_{\mu \nu} &=& \na_{[ \mu }{\tilde B}_{\nu ]}.
\een
It turns out in section 4, that in order to envisage the influence of {\it dark matter} sector on the Hall viscosity, we should take into account the effects of {\it dark matter}
backreaction on the metric. To do this we expand all the adequate quantities in series in $\alpha $-coupling constant and calculate the backreaction up to the linear order.

Therefore,  in what follows we shall use the action (\ref{s_matter}), where one has the explicit dependence on $\alpha$-coupling constant
The equations of motion  obtained from the variation of the action $S=S_g+S_m$ with respect
to the metric, the scalar field and gauge fields are given by
\ben \label{eq1}
G_{\mu \nu} &+& \Lambda~g_{\mu \nu} - \la~C_{\mu \nu} = T_{\mu \nu}(\theta) + T_{\mu \nu}(F) + T_{\mu \nu}(B) 
+ \alpha~T_{\mu \nu}(F,~B),\\
\na_\mu \na^\mu \theta &-& \frac{\p V}{\p \theta} = \frac{\la}{4}~{}^{\ast}R~R,\\
\na_{\mu}F^{\mu \nu} &+& \frac{\alpha}{2}~\na_\mu B^{\mu \nu} = 0,\\ \label{eq2}
\na_{\mu}B^{\mu \nu} &+& \frac{\alpha}{2}~\na_\mu F^{\mu \nu} = 0.
\een
The contributions to the energy momentum tensors are given by 
\ben
T_{\mu \nu} (\theta) &=& \frac{1}{2} \na_{\mu} \theta \na_\nu \theta - \frac{1}{4}~g_{\mu \nu}~\na_\delta \theta \na^\delta \theta - \frac{1}{2}~g_{\mu \nu} ~V(\theta),\\
T_{\mu \nu}(F) &=& \frac{1}{2}~F_{\mu \delta}F_{\nu}{}^{\delta} - \frac{1}{8}~g_{\mu \nu}~F_{\alpha \beta}F^{\alpha \beta},\\
T_{\mu \nu}(B) &=& \frac{1}{2}~B_{\mu \delta}B_{\nu}{}^{\delta} - \frac{1}{8}~g_{\mu \nu}~B_{\alpha \beta}B^{\alpha \beta},\\
T_{\mu \nu}(F,~B) &=& \frac{1}{2}~F_{\mu \delta}B_{\nu}{}^{\delta} - \frac{1}{8}~g_{\mu \nu}~F_{\alpha \beta}B^{\alpha \beta}.
\een
The above system of equations can be reduced to the following relations:
\be
R_{\mu \nu} + 3~g_{\mu \nu} - \la~C_{\mu \nu} = t_{\mu \nu}(\theta) + T_{\mu \nu}(F)  +  T_{\mu \nu}(B) 
+ \alpha~ T_{\mu \nu}(F,~B) ,
\ee
where we have denoted
\ben
t_{\mu \nu}(\theta) &=& \frac{1}{2}~\na_{\mu} \theta \na_{\nu} \theta + \frac{1}{2}~g_{\mu \nu}~V(\theta),\\
C^{\mu \nu} &=& \na_\gamma \theta~\ep^{\gamma \kappa \delta (\mu}~\na_{\delta}R_{\kappa}{}^{\nu)} + \na_\gamma \na_\delta \theta~{}^{\ast}R^{\delta (\mu \nu) \gamma}.
\een

In order to find the Hall viscosity we have to compute its contribution to the hydrodynamics flow of the boundary field theory.
Therefore, it is important to write the stress tensor connected with the action of the considered theory, as we have to keep the fields fixed at the boundary.
The general procedure is similar to that of finding the boundary Gibbons-Hawking term in general relativity. Namely, the variation of the boundary term spoils the principle
leading to Einstein equations because of the fact that it contains the contribution proportional to the extrinsic curvature of the boundary. Thus, the variation will constitute
two terms, a bulk piece that vanishes when the equations of motion are fulfilled and a boundary term. The situation can be cured by adding to the action a counter term
canceling the boundary one.

In the case of Chern-Simons theory,
it can be shown that if $\theta$ pseudo scalar field vanishes asymptotically for any solutions of the equations of motion derived from the action in the theory in question,
the stress energy tensor is the same as of an asymptotically $AdS_4$ spacetime. Namely, taking variations of the Chern-Simons gravity action we obtain \cite{sar12, kra06,gru08}
\be
\delta S_{CS} = - \frac{\la}{4}~\int d^4x~\sqrt{-g}~\theta~{}^\ast R~R= -\la~\delta S_1 -\la~\delta S_2  + \la~\delta S_3,
\ee
where the explicit forms are given by
\ben
 \delta S_1 &=& \int d^4x~\sqrt{-g}~\delta g_{\alpha \beta}~\na_\la \na_\gamma ( \theta~{}^\ast R^{\la \alpha \gamma \beta}),\\
 \delta S_2  &=&  \int d^4x~\sqrt{-g}~\na_\alpha (\theta {}^\ast R^{\beta~\gamma \la}_{~\xi}~\delta \Gamma^{\xi}_{\beta \la}),\\
\delta S_3 &=& \int d^4x~\sqrt{-g}~\na_\beta \bigg[ \delta g_{\la \gamma}~\na_{\xi}(\theta {}^\ast R^{\beta \la \xi \gamma}) \bigg].
\een
It can be proved that
\be
\delta S_1 = \int d^4x~\sqrt{-g}~\delta g_{\alpha \beta} ~C^{\alpha \beta},
\ee
whereas the term $\delta S_2 $ can be cast in the form of a variation of the extrinsic curvature plus the term containing dual curvature term multiplied by the variation of the second kind of Christofel
symbol. Using the Gaussian normal coordinates 
\be
ds^2 = d\eta^2 + g_{ij} dx^i dx^j,
\ee
and the Codazzi equation, 
as well as having in mind the fact that in the spacetime with negative cosmological constant, solutions of Einstein equations admit the expansion (the so-called Fefferman-Graham expansion \cite{fef85}) provided by
\be
g_{ij} = e^{2\eta} ~ g_{ij}^{(0)} + g_{ij}^{(2)} + e^{-2 \eta} ~g_{ij}^{(4)} + \dots
\ee
one may find that the term under consideration, i.e.  $\delta S_2$, vanishes as
$\int_{\p M} d^3x~\theta \sim 0$.  We can think about the boundary as being situated at $\eta \rightarrow \infty$, with the metric tensor conformal to $g_{ij}^{(0)}$.

In the case of $\delta S_3$, with a help of Bianchi indentity, it can be envisaged that its boundary behavior is of the form
$\ep^{\alpha \beta \gamma \la}~\na_\alpha \theta~\na_\gamma K^{\xi}_{\la}~\delta g_{\xi \beta} \rightarrow 0$. Thus, the only relevant term to the variation of the action is
the conventional Gibbons-Hawking one. On the other hand, the only counter term will be a boundary cosmological constant renormalization, because of the fact that the boundary under consideration is flat.
Then, the stress tensor will be provided by
\be
\delta S = \frac{1}{2} \int d^3x~\sqrt{g_{ij}^{(0)}}~T^{ij}~\delta g_{ij}^{(0)},
\ee
where $g_{ij}^{(0)}$ is the metric on the conformal boundary.

\section{Equations of motion for the perturbed system}
\label{sec:eqs}
In this section we shall consider the boosted black brane solution in Einstein-Maxwell-{\it dark matter} gravity in AdS four-dimensional spacetime.
To derive the hydrodynamic equations and response parameters we have to perturb the velocity field, black brane 
temperature and charge in the bulk. Such perturbations back-react on the metric and thus change the background.
We shall calculate the back-reaction perturbatively by expanding the relevant functions up to the linear order
in derivatives. In the present theory these functions additionally depend on the coupling to the dark matter. To capture
this dependence analytically we shall assume small value of $\alpha$ and expand all relevant functions with respect to it.
Thus we are dealing with two expansions and in this paper we limit the calculations to the linear order 
in the coupling to the {\it dark matter} sector and the perturbing fields.
 
In order to solve the equations of motion perturbatively, order by order, 
we write the line element expanded up to the first order 
in the boundary derivatives around the coordinates origin, $x_\nu = 0$. Having obtained the 
metric solution near the origin one can extend it to the whole manifold iteratively
\cite{bha08,raa08}.

It turns out that the line element satisfying the equations of motion  can be written as
\ben 
ds^2 &=& -2H(r,T,Q)u_\alpha dx^\alpha dr - r^2F(r,T,Q)u_\alpha u_\beta dx^\alpha dx^\beta  + r^2(\eta_{\alpha \beta} + u_\alpha u_\beta )dx^\alpha dx^\beta ,\\
A &=& A(r,~T,~Q)u_\alpha dx^\alpha,\\ 
B &=& B(r,~T,~Q)u_\alpha dx^\alpha,\\ 
\theta &=& \theta(r,~T,~Q).
\een
By $T$ and $Q$, we have denoted the Hawking temperature and the charge of the boosted black brane.
This background geometry describes hydrodynamics in $(2+1)$-dimensional spacetime 
at thermal equilibrium at the boundary. The velocity is given by
\be
u^\nu = \frac{1}{\sqrt{1-\beta^2}}(1,~\beta^i).
\ee
If one allows the constant quantities like $T,~Q,~A_\mu,~B_\mu,~\theta$ to  be slowly varying functions 
of the boundary coordinates than the expansions near the coordinate origin, given to the first order,  provided by
\ben
F(r, T, Q) &=& F(r) + \frac{\p F}{\p T} x^\mu \p_\mu T + \frac{\p F}{\p Q} x^\mu \p_\mu Q = F(r) + \delta F,\\
H(r,T,Q) &=& Q(r) + \frac{\p H}{\p T} x^\mu \p_\mu T +  \frac{\p H}{\p Q} x^\mu \p_\mu Q = H(r) + \delta H,\\
A(r,Q,T) &=& A(r) + \frac{\p A}{\p T} x^\mu \p_\mu T +  \frac{\p A}{\p Q} x^\mu \p_\mu Q = A(r) + \delta A,\\
B(r,Q,T) &=& B(r) + \frac{\p B}{\p T} x^\mu \p_\mu T +  \frac{\p B}{\p Q} x^\mu \p_\mu Q = B(r) + \delta B,\\
\theta(r,Q,T) &=& \theta(r) + \frac{\p \theta}{\p T} x^\mu \p_\mu T +  \frac{\p \theta}{\p Q} x^\mu \p_\mu Q = \theta(r) + \delta \theta,\\
u^\alpha &=& (1,~x^\ga \p_\ga \beta^i),\\
T &=& T_0 + x^\mu \p_\mu T, \\
Q &=& Q_0 + x^\mu \p_\mu Q,
\een
are sufficient to calculate the viscosities. As we have already mentioned, the above functions do depend on $\alpha$, but this 
will be discussed later on. In agreement with previous studies~\cite{bri08,sar12,che12,roy14} we  
perform the entire analysis in the comoving frame, where the fluid velocity equals zero at the boundary. 
The resulting inhomogenous background line element and the gauge fields yield
\ben \label{an1}
ds^2 &=& 2~H(r)~dv~dr - r^2~F(r)~dv^2 + r^2~dx^\mu~dx_\mu + \\ \nonumber
&+& \ep \bigg[
2~\delta H~dv~dr - 2~H(r) x^\alpha \p_\alpha \beta_\ga dx^\ga dr - r^2~\delta F~dv^2 + 2~r^2 (F(r)-1)x^\mu \p_\mu \beta_\ga ~dv~dx^\ga
\bigg], \\
\theta &=& \theta(r) + \ep~\delta \theta,\\
A &=& -A(r)~dv + \ep~\bigg[ A(r)~x^\ga \p_\ga \beta_\zeta ~dx^\zeta \bigg],\\
B &=& -B(r)~dv + \ep~\bigg[ B(r)~x^\ga \p_\ga \beta_\zeta ~dx^\zeta \bigg],
\een
where the small parameter $\ep$ serves as bookkeeping device, which power denotes the number of the 
derivatives taken into account along the boundary.

It is important to notice, that with parameters depending on the boundary coordinates, the ansatz (\ref{an1}) 
of the background line element does not satisfy the equations of motion for the underlying theory.
Thus the next step is to correct the  line element order by order in $\ep$, 
to make the metric tensor, the gauge fields and the pseudo scalar $\theta$ fulfill 
equations (\ref{eq1})-(\ref{eq2}), order by order
in the perturbation series. The correction to the line element is taken in  the following form
\ben
ds^{(1) 2}_{corr} &=& \ep \bigg[
\frac{k(r)}{r^2}dv^2 + 2~p(r)~dvdr - r^2~p(r)~dx_i dx^i + \frac{2}{r}~w_i(r)~dv dx^i + \\ \nonumber
&+&  r^2~\alpha_{ij}(r)~dx^i dx^j \bigg],\\
\theta_{corr} &=& \ep~\theta,\\
A_{corr} &=& \ep~({\tilde a}_v(r) dv + {\tilde a}_m(r) dx^m),\\
B_{corr} &=& \ep~({\tilde b}_v(r) dv + {\tilde b}_m(r) dx^m),
\een
where $\alpha_{ij}$ is symmetric and traceless. 

\subsection{Zeroth order equations}

From the equation (\ref{an1}) we read off the zeroth order line  element which is provided by
\be
ds^2 = 2~H(r)~dv~dr - r^2~F(r)~dv^2 + r^2~(dx^2 + dy^2),
\ee
and find the following equations of motion: 
\ben
F^{''}(r)  &+& \left(\frac{6}{r}-~\frac{H'(r)}{H(r)}\right)~F'(r) + \frac{2}{r}\left(\frac{3}{r}-\frac{H'(r)}{H(r)}\right)F(r)~
\\ \nonumber
&-& \frac{1}{r^2}\left(6-V(\theta)\right)H^2(r) 
= \frac{A'(r)^2}{2~r^2} + \frac{B'(r)^2}{2~r^2} + \alpha~\frac{A'(r)~B'(r)}{r^2},\\ \label{xxc}
F'(r) &+&\left(\frac{3}{r}-\frac{H'(r)}{H(r)}\right)F(r) - \frac{1}{2r}[6- V(\theta)] H^2(r) +\\ \nonumber
&+& \frac{A'(r)^2}{4~r} + \frac{B'(r)^2}{4~r} + \alpha~\frac{A'(r)~B'(r)}{4~r} = 0,\\
\frac{H'(r)}{H(r)} &=& \frac{r}{4}~\theta'(r)^2,\\
A^{''}(r)  + A'(r)~\bigg( \frac{2}{r} &-& \frac{H'(r)}{H(r)} \bigg) + \frac{\alpha}{2}~\bigg[
B'(r)~\bigg( \frac{2}{r} - \frac{H'(r)}{H(r)} \bigg) + B^{''}(r) \bigg] = 0,
\label{e326}\\
B^{''}(r)  + B'(r)~\bigg( \frac{2}{r} &-& \frac{H'(r)}{H(r)} \bigg) + \frac{\alpha}{2}~\bigg[
A'(r)~\bigg( \frac{2}{r} - \frac{H'(r)}{H(r)} \bigg) + A^{''}(r) \bigg] = 0,
\label{e327} \\
\theta^{''}(r) + \theta'(r)~\bigg(
\frac{4}{r} + \frac{F'(r)}{F(r)} &-& \frac{H'(r)}{H(r)} \bigg) - \frac{H^2(r)}{r^2~F(r)}~\frac{\p V}{\p \theta} = 0,
\een
where the prime denotes  derivation with respect to $r$-coordinate.


Having in mind equation for $\theta$ field, one remarks that its asymptotic behavior  
is of the form
\be
\theta = \frac{\cO_-}{r^{\Delta_-} }+ \frac{\cO_+}{r^{\Delta_+} }+ \dots,
\ee
where $\Delta_{\pm} = 3/2 \pm \sqrt{9/4 + m^2}$. In the following, 
we turn off the mode $\cO_-$ and the mode $\cO_+$ is identified with the condensate at the boundary.
It turns out that \cite{kle99} for $-9/4 <m^2<-5/2$, both $\cO_-$ and $\cO_+$ are renormalizable and one can point either one as a source and the other as a condensate.

In the neutral Hall viscosity 
case \cite{sar12}, $\cO_-$ can be turned off only if 
$c<-\frac{3}{4}$, which in turn violates the positive energy condition \cite{her06} 
and makes the solution in question unstable.
In the considered case of $V(\theta)$ being a Mexican hat type potential, the $\theta^4$ term engenders that the 
solution is regular at the event horizon of charged black brane.

 To proceed further let us note that the symmetry of equations (\ref{e326}) and (\ref{e327}) is such that
both $A(r)$ and $B(r)$ independently of each other have to fulfill the equations (\ref{e331}) and (\ref{e332}) below.
For future convenience we rewrite the whole set of equations  as
\ben \label{e330}
F'(r) &+& F(r)~C(r) + D(r) + E(\alpha,~r) = 0,\\
A'' (r) &+& A'(r)~\bigg( \frac{2}{r} - \frac{r}{4}~\theta'(r)^2 \bigg)= 0,
\label{e331} \\
B'' (r) &+& B'(r)~\bigg( \frac{2}{r} - \frac{r}{4}~\theta'(r)^2 \bigg)= 0,
\label{e332} \\
\theta^{''}(r) &+& \theta'(r)~\bigg(
\frac{4}{r} + \frac{F'(r)}{F(r)} \bigg) - \frac{r}{4 } \theta'(r)^3  - \frac{H^2(r)}{r^2~F(r)}~\frac{\p V}{\p \theta} = 0, \\  \label{e334}
\frac{H'(r)}{H(r)} &=& \frac{r}{4}~\theta'(r)^2,
\een
where we have set
\ben
C(r) &=& \frac{3}{r} - \frac{r}{4}~\theta'(r)^2,\\
D(r) &=& - 3~\frac{H^2(r)}{r} + \frac{V(\theta)~H^2(r)}{2~r} + \frac{A'(r)^2}{4~r} + \frac{B'(r)^2}{4~r} ,\\ \label{e337}
E(\alpha,~r) &=&  \alpha~\frac{A'(r)~B'(r)}{4~r}.
\een
We are interested in the effect of dark matter on the viscosities. Even though $\alpha$ enters 
equations only {\it via} $E(\alpha,r)$ above, other functions implicitly depend on it. To access this
dependence in an analytic form we expand all functions to  the lowest, linear order
\ben \label{e338}
F(r) &=& F^{(0)}(r) + \alpha~f(r) + \cO(\alpha^{n \geq 2}),\\
H(r) &=& H^{(0)}(r) + \alpha~h(r) + \cO(\alpha^{n \geq 2}),\\
A(r) &=& A^{(0)}(r) + \alpha~a(r) + \cO(\alpha^{n \geq 2}),\\
B(r) &=& B^{(0)}(r) + \alpha~b(r) + \cO(\alpha^{n \geq 2}),\\ \label{e342}
\theta(r) &=& \theta^{(0)}(r) + \alpha~\zeta(r) + \cO(\alpha^{n \geq 2}).
\een
 We note in passing that to expect all functions to depend in a linear manner on $\alpha$ is not
obvious, but it is natural to expect linear dependence due to the fact that only the first power of $\alpha$ 
enters the system of equations to be solved.

The $\alpha^0$-order equations are given by
\ben 
F^{(0) '}(r)  &+& F^{(0)}(r)~C^{(0)}(r) + D^{(0)}(r) = 0,\\
\frac{H^{(0)'}(r)}{H^{(0)}(r)} &=& \frac{r}{4}~{\theta^{(0)' 2}(r)},\\
A^{(0)'' }(r) &+& A^{(0)'}(r)~\bigg( \frac{2}{r} - \frac{r}{4}~\theta^{(0)'}(r)^2 \bigg)= 0,\\
B^{(0)''} (r) &+& B^{(0)'}(r)~\bigg( \frac{2}{r} - \frac{r}{4}~\theta^{(0)'}(r)^2 \bigg)= 0,\\ 
\theta^{(0)''}(r) &+& \theta^{(0)'}(r)~\bigg(
\frac{4}{r} + \frac{F^{(0)'}(r)}{F^{(0)}(r)} \bigg) - \frac{r}{4 } \theta^{(0)' 3}(r)  + \\
&-&
\bigg( m^2~\theta^{(0)}(r) + c~\theta^{(0)3}\bigg)~\frac{H^{(0)2}(r)}{r^2~F^{(0)}(r)} =0,
\een

The inspection of the above relations enables us to solve them as the first order differential equation.  In zeroth order we obtain the following:
\ben
H^{(0)}(r) &=& h_1~exp\bigg[ \int_r^\infty ds~\frac{s}{4}~\theta^{(0)' 2}(s) \bigg],\\
F^{(0)}(r) = \bigg[ &-& \int_r^\infty dr~ D^{(0)}(r)~ exp \bigg( \int_x^\infty dx~C^{(0)}(x) \bigg) + C_2 \bigg]  \times \\ \nonumber
\times ~exp \bigg[ &-& \int_r^\infty ds~ C^{(0)}(s) \bigg],
\een
where $C_2$ and $h_1$  are constants.

In the next step we find the equations in $\alpha^1$-order. They are given by
\ben \label{e351}
f'(r) &+& f(r)~\bigg( \frac{3}{r} - \frac{r}{4}~\theta^{'(0) 2}(r) \bigg)+ Q(r) = 0,\\
a''(r) &+& a'(r)~\bigg( \frac{2}{r} - \frac{r}{4}~\theta^{'(0) 2}(r)  \bigg) - \frac{r}{2}~\zeta'(r)~\theta^{'(0)}(r)~A^{'(0)}(r) = 0,\\
b''(r) &+& b'(r)~\bigg( \frac{2}{r} - \frac{r}{4}~\theta^{'(0) 2}(r)  \bigg) - \frac{r}{2}~\zeta'(r)~\theta^{'(0)}(r)~B^{'(0)}(r) = 0,\\
h'(r) &-& \frac{H^{'(0)}(r)}{H^{(0)}(r)}~h(r) - r~\zeta'(r)~\theta^{'(0)}(r)~H^{(0)}(r) = 0,\\ \label{e355}
\zeta''(r) &+& \zeta'(r)~\bigg( - r~\theta^{'(0) 2}(r) + \frac{4}{r} + \frac{F^{' (0)}(r)}{F^{(0)}(r)} \bigg)
+ \zeta(r)~M(r) + K(r) = 0,
\een
where $Q(r),~M(r)$ and $K(r)$ are denoted by
\ben
Q(r) &=& - \frac{r}{2}F^{(0)}(r) \zeta'(r) \theta^{'(0)}(r) - \frac{6}{r}h(r) H^{(0)}(r) + \frac{a'(r) A^{'(0)}(r)}{2r}  + \frac{b'(r) B^{'(0)}(r)}{2r} \\ \nonumber
&+& \frac{m^2}{2} \bigg[ 2 \theta^{(0) 2}(r) h(r) H^{(0)}(r) + \frac{\zeta(r) \theta^{(0)}(r) H^{(0)2}(r)}{r} \bigg] 
+ \frac{A^{'(0)}(r)  B^{'(0)}(r)}{4 r}
\\ \nonumber
&+& \frac{c}{2} \bigg[
\frac{\theta^{(0)3}(r) \zeta(r) }{r} + 2 h(r) H^{(0)}(r) \theta^{(0) 4} \bigg],\\
M(r) &=& - \frac{ r}{2} \theta^{(0)'2}(r) - \bigg( m^2 + \theta^{(0)2}(r) \bigg) ~\frac{H^{(0)2}(r)}{r^2~F^{(0)}(r)},\\ \label{e358}
K(r) &=& - \bigg( m^2 \theta^{(0)}(r) + c \theta^{(0)3} \bigg) \bigg(
\frac{2 H^{(0)}(r) h(r)}{r^2 F^{(0)}(r)} - \frac{f(r) H^{(0)2}(r)}{r^2 F^{(0)2}(r)} \bigg) \\ \nonumber
&+& \bigg( \frac{f'(r)}{F^{(0)}(r)} - \frac{F^{' (0)}(r) f(r)}{F^{(0)}(r)} \bigg) \theta^{(0)'}.
\een
As in the previous  zeroth-order case, we have obtained two linear differential equations bounded with the fluctuations of the underlying line element. The exact forms
of the $\alpha^1$-corrections of $F(r)$ and $H(r)$ are given by the relations 
\ben
f(r) = \bigg[ C &-& \int_r^\infty dr~Q(r)~exp\bigg[ \int_{r'}^\infty dr'~\bigg(\frac{3}{r'} - \frac{r'}{4}~\theta^{'(0) 2}(r')\bigg)\bigg] \times \\
\times exp\bigg[ &-& \int_x^\infty dx~\bigg(\frac{3}{x} - \frac{x}{4}~\theta^{'(0) 2}(x)\bigg)\bigg] ,\\
h(r) = \bigg[ D &+& \int_r^\infty dr~ (r~\zeta'(r)~\theta^{'(0)}(r)~H^{(0)}(r) )~exp\bigg[ - \int_x^\infty dx \frac{H^{'(0)}(x)}{h^{(0)}(x)} \bigg] \bigg] \times \\ \nonumber
 \times exp\bigg[ &-& \int_s^\infty ds \frac{H^{'(0)}(s)}{h^{(0)}(s)} \bigg]. 
\een
The important question  concerns the validity of $\alpha$-coupling constant expansion. 
At the first step let us consider the situation when $\alpha = 0$.
From the relation (\ref{e327}) one gets that $B$ is equal to zero if this {\it dark matter} 
field does not carry the chemical potential. On the other hand, the scalar field
$\theta$ can be non-zero due to the fact that it is responsible for the instabilities.

The other case to consider is connected with the fact when $\alpha$-coupling constant is small,
 equations (\ref{e351})-(\ref{e358}). As {\it dark matter} gauge field component is equal to zero
for the background field, then the leading correction for the other quantities in question 
can be set to zero. In the process of this, one receives
the homogeneous system of differential equations (it seems that $Q(r)$ destroys homogeneity 
but $B^{(0)}=0$ and it is equal to zero). The conclusion one can draw is that nothing sources
the perturbative correction. Moreover, by similar arguments applied to the exact
set of equations (\ref{e330})-(\ref{e337}) one can show that corrections 
 to all orders in $\alpha$ vanish unless $\mu_D \ne 0$.

Therefore in what follows, in order to obtain the expansions described by the relations (\ref{e338})-(\ref{e342}), we assume that the {\it dark matter} gauge field carries the 
deformation chemical potential, i.e., $B (r \rightarrow \infty) = \mu_D$. This is a quite new situation comparing to the description of Hall viscosity in the AdS Einsten-Maxwell case \cite{che12}.

\section{The viscosities in the presence of dark matter sector}
\label{sec:hall}
In this section we shall compute the shear and Hall viscosities in the theory under consideration. As was justified in \cite{bha08, raa08} viscosity components can be 
found by the inspection of the spatial components of the energy-momentum tensor like
$T_{xy}$ or $T_{xx} - T_{yy}$.
To commence with let us consider the following:
\be
R_{xy}^{(1)} + 3~g_{xy}^{(1)} - \la~C_{xy}^{(1)} = t_{xy}^{(1)}(\theta) + T_{xy}^{(1)}(F) + T_{xy}^{(1)}(B) + \alpha~T_{xy}^{(1)}(F,~B),
\ee
where we have denoted by the superscript $(1)$ the fluctuations connected with the leading order $\cO(\ep)$ in the derivative expansion.
The exact form of the above relation is provided by 
\ben \label{all}
\frac{1}{H(r)}\frac{d}{dr} \bigg[ &-& \frac{1}{2}~\frac{r^4~F(r)}{H(r)}~\frac{d \alpha_{xy}}{dr} \bigg]
+ \bigg(
\frac{r^3~H'(r)~F(r)}{H^3(r)} - \frac{r^3~F'(r)}{H^2(r)}  - 3 ~\frac{r^2~F(r)}{H^2(r)}  \\ \nonumber
&+& 3r^2 - \frac{r^2}{2}~V(\theta) - \frac{r^2~A'(r)^2}{4~H^2(r)} - \frac{r^2~B'(r)^2}{4~H^2(r)}  - \alpha~\frac{r^2~A'(r)~B'(r)}{4~H^2(r)} 
\bigg)~\alpha_{xy} \\ \nonumber
&=& \frac{r}{H(r)}~(\p_x \beta_y + \p_y \beta_x) + \frac{\la}{4~H(r)} ~\frac{d}{dr}
\bigg(
\frac{r^4~F'(r)~\theta'(r)}{H^2(r)} \bigg)~(\p_x \beta_y - \p_y \beta_x).
\een
One can see that having in mind equation (\ref{xxc}), the second term in the relation (\ref{all}) is exactly equal to $-r^3/H^2(r)$ multiplied by (\ref{xxc}) and hence equal to zero.
It implies that for $\alpha_{xy}$ we obtain
\be
\alpha_{xy} = \int_r^\infty dl~\frac{2~H(l)}{l^4~F(l)}~\int_{r_{H}}^r ds~\bigg[ s~(\p_x \beta_y + \p_y \beta_x) + \frac{\la}{4}~\frac{d}{ds} \bigg(
\frac{s^4~F'(s)~\theta'(s)}{H^2(s)} \bigg)~(\p_x \beta_x - \p_y \beta_y)
\bigg].
\ee
In order to compute the asymptotical form of $\alpha_{xy}$ we shall implement the formula \cite{sar12} which is valid as $r \rightarrow \infty$, namely
\be
r^n~\alpha_{xy} (r)  \rightarrow - \frac{r^{n+1}}{n}~\frac{d}{dr} \alpha_{xy}(r).
\ee
Using the Graham-Fefferman coordinate system it was shown \cite{har01} that up to six dimensions of the spacetime, the expectation value for the stress-energy
tensor of the dual theory is provided by
\be
<T_{ij}> = \frac{n }{16 \pi~G_N} g_{ij (n)} + X_{ij}(g_{(n)}),
\ee
where $n$ denotes the spacetime dimension and $X_{ij}(g_{(n)})$ is a function of metric tensor components 
\be
g_{ij}(x_\mu,~r) = g_{ij (0)} + \frac{1}{r^2} g_{ij (2)} + \dots + \frac{1}{r^n} g_{ij (n)} + \dots
\ee
It can be seen that the exact form of $<T_{ij}>$ depends on the dimensionality 
of the spacetime in question and indicates the conformal
anomalies of the boundary conformal field theory. In the odd dimensional spacetime, when 
the gravitational conformal anomalies are equal to zero,
it reduces to 
\be
<T_{ij}> = \frac{n }{16 \pi~G_N} g_{ij (n)} .
\ee
One remarks also that when $r \rightarrow \infty$ the values of the functions $F(r)$ and $H(r)$ tend to $1$.
All the above reveal that in the case under consideration we get
\ben
<T_{xy}> = \frac{3 }{16 \pi~G_N} \alpha_{xy (3)} &=& - \frac{1}{16 \pi G_N}~(\p_x \beta_y + \p_y \beta_x) \\ \nonumber
&-& \frac{1}{8 \pi G_N}~\bigg[ (\p_x \beta_y - \p_y \beta_y)~\frac{\la}{4}~\frac{r^4~F'(r)~\theta'(r)}{H^2(r)} \bigg]_{r= r_H}.
\een
The first term is the usual shear mode with the shear viscosity $\eta_s$ provided by
\be
\eta_s = \frac{1}{16 \pi G_N}.
\ee
It can be seen that $\eta_s$ is not corrected by the $\alpha$-coupling constant, at the leading order in it,
while the second one is proportional to the Hall viscosity 
\be
\eta_H=  - \frac{1}{8 \pi G_N}~\frac{\la}{4}~\frac{r^4~F'(r)~\theta'(r)}{H^2(r)} \mid_{r=r_H}.
\ee
When one divides it by the entropy density, we receive the Hall viscosity/ entropy ratio complement 
to Einstein-Maxwell {\it dark matter} Chern-Simons theory in $AdS_4$ spacetime
subject to the backreaction effects. It depends linearly an the coupling $\alpha$
do dark matter and reads 
\be
\frac{\eta_H}{s} = \frac{\eta^{(0)}_H}{s}~\bigg( 1 + \alpha~\Sigma \bigg) + \cO(\alpha^{n\geq2}),
\ee
where we set for $\eta_H^{(0)}/s$ and $\Sigma$ the following relations:
\ben
\frac{\eta^{(0)}_H}{s} &=& - \frac{\la}{2~\pi}\bigg( \frac{r^4~F^{' (0)}(r)~\theta^{' (0)}(r)}{H^{2 (0)}(r)} \bigg)_{\mid_{r=r_H}},\\
\Sigma &=& \bigg( \frac{f'(r)}{F^{' (0)}(r)} + \frac{\zeta'(r)}{\theta^{' (0)}(r)} - \frac{2~h(r)}{H^{(0)}(r)} \bigg)_{\mid_{r=r_H}}.
\een
The term $\Sigma$ is a direct consequence of the presence of the backreaction of the matter on the metric.
The sign of the correction is not uniquely determined.

\subsection{Dependence of Hall viscosity on condensation value }
As noted earlier the shear viscosity $\eta_s$ takes on universal value not 
modified by the presence of {\it dark matter} or condensation value. On the contrary, the condensation 
of the pseudo scalar field $\theta$ is essential to get the parity breaking~\cite{che12} and consequently 
non-zero value of the Hall viscosity. Here we analyze  the dependence of the Hall viscosity 
on the condensation value  and temperature close to the critical one, $T_c$.

Let us consider the explicit form of the charged black brane line element \cite{iqb10}
\be 
ds^2 = 2~dv~dr - r^2~F(r)dv^2 + r^2~(dx^2 + dy^2),
\ee
where the component of the metric tensor $F(r)$ and the gauge field are provided by
\be
F(r) = 1 - \frac{1 + 3 \kappa}{r^3} + \frac{3 \kappa}{r^4}, \qquad A= 2~\sqrt{3~\kappa}\bigg( 1 - \frac{1}{r} \bigg)~dv.
\ee
In what follows we set that the event horizon is situated at $r_H =1$.
In the chosen units the Hawking temperature and the chemical potential are given, respectively as
\be
T_{BH} = \frac{3}{4 \pi}~(1 - \kappa), \qquad \mu = 2~\sqrt{3~\kappa}.
\ee
We shall use $z=1/r$ coordinates in which the equation of motion for $\theta$ field implies
\be
\theta'' + \theta'~\bigg( \frac{F'(z)}{F(z)} - \frac{2}{z} \bigg) - \frac{m^2~\theta + c~\theta^3}{z^2~F(z)} = 0.
\label{thetabis}
\ee
To proceed further, let us expand $\theta(z)$ near the black brane event horizon where $z=1$. It yields
\be
\theta(z) = \theta(1) + \theta'(1)~(z-1) + \frac{1}{2}\theta'' (1)~(z-1)^2 + \dots, 
\label{expans}
\ee
with finite values of $\theta'(1)$ and $\theta''(1)$. Note also that with our choice of units $F(1)=0$. 
Calculating the limit $z\rightarrow 1$ in equation (\ref{thetabis}) we get  first the relation for $\theta'(1)$
\be
\theta'(1) = \frac{m^2~\theta(1) + c~\theta^3(1)}{F'(1)},
\ee
and $\theta''(1)$
\be
\theta''(1) = \frac{\theta'(1)}{2}~\bigg[
\frac{~( m^2 + 3~c~\theta^2(1) - F^{''}(1))}{F'(1)} \bigg].
\ee
Using the above relations, we find the approximate form of $\theta(z)$ near the black brane event horizon 
\ben
\theta(z) = \theta(1) &+& (z-1)~\frac{m^2~\theta(1) + c~\theta^3(1)}{F'(1)} + \\ \nonumber
&+&  \frac{(z-1)^2}{4}~\bigg[
\frac{\theta'(1)~( m^2 + 3~c~\theta^2(1) - F^{''}(1))}{F'(1)} \bigg] + \dots
\een
On the other hand, in the asymptotic AdS region, when $z \rightarrow 0$, ~$\theta$ behaves like $\cO_+~z^{\Delta_+}$.
In order to find $\theta(1)$ and $\cO_+$, we match smoothly the solution at the event horizon and in AdS region, in some intermediate point $z_m$
\be
\theta_{H} (z_m) = \theta_{boundary}(z_m), \qquad \theta'_{H} (z_m) = \theta'_{boundary}(z_m).
\ee
Namely, one arrives at the following:
\ben
\cO_+~z_m^{\Delta_+} &=& \theta(1) + (z_m-1)~\frac{m^2~\theta(1) + c~\theta^3(1)}{F'(1)} + \\ \nonumber
&+&  \frac{(z_m-1)^2}{4}~\bigg[
\frac{\theta'(1)~( m^2 + 3~c~\theta^2(1) - F^{''}(1))}{F'(1)} \bigg] ,\\
\cO_+~\Delta_+~z_m^{\Delta_+ -1} &=& ~\frac{m^2~\theta(1) + c~\theta^3(1)}{F'(1)} + \\ \nonumber
&+& \frac{(z_m-1)}{2}~\bigg[
\frac{\theta'(1)~( m^2 + 3~c~\theta^2(1) - F^{''}(1))}{F'(1)} \bigg] .
\een
We can draw a conclusion that $\theta(1) \approx \cO_+$. On the other hand, as was revealed in \cite{nak14,nak15,nak15a1,rog15,rog15a}, the condensation value is proportional to
$\sqrt{1 - \frac{T}{T_c}}$. 
It suggests that at the critical temperature $T_c$, the Hall viscosity is independent on $\alpha$-coupling constant, i.e., 
$\eta_H$ does not depend on the {\it dark matter} sector, in the probe limit. As it has been shown in the preceding sections
 only the backreaction  effects are responsible for revealing the aforementioned dependence.

\section{Summary and conclusions}
\label{sec:conclusions}
 We have studied the holographic fluid in a model containing Maxwell gauge field and
the other $U(1)$ field describing the dark matter with the goal to analyze the influence of the latter
on the fluid viscosities. To study two-dimensional flow we have 
used $(3+1)$-dimensional bulk and pseudo scalar field $\theta$ coupled to the gravitational 
Chern-Simons term.  The arena for our investigations is the anti de Sitter spacetime of charged black 
brane of finite temperature. The bulk pseudo scalar potential is composed of $\theta^2$ and 
$\theta^4$ terms which guarantee  consistent solution at zero temperature. 
Contrary to the previous studies of Hall viscosity in AdS Einstein-Maxwell case \cite{che12}, the spontaneously 
broken parity by the pseudo scalar hair, as well as, the {\it dark matter} gauge field deformation chemical potential,
give rise to the emergence of a non-zero value of the Hall 
viscosity at the boundary.

We have solved the underlying equations of motion perturbatively up to the leading order in the  
$\alpha$-coupling constant and found that the shear viscosity mode
is not corrected at the leading order by the presence of  {\it dark matter}.
 On the contrary, the correction of the Hall viscosity to entropy ratio is modified linearly in $\alpha$.
The leading term  is the same  as earlier derived in the model without dark matter~\cite{sar12,che12}.

Parity or time reversal symmetry breaking is a necessary condition for the
existence of Hall viscosity in the system with gauge fields. In the present approach the parity 
is broken spontaneously by the condensation of the $\theta$ field below certain transition temperature $T_c$. 
We have found that the Hall viscosity depends on temperature as  $\eta_H \sim \sqrt{1 - \frac{T}{T_c}}$.
It is important to note that neither prefactor nor transition temperature itself 
in the  above relation,  depend on the coupling constant $\alpha$ of the {\it dark matter}. This  makes the 
condensation of $\theta$ field completely different from earlier studied superconductors~\cite{nak14,nak15,nak15a1,rog15,rog15a}.
In the latter case both the condensation value and $T_c$ showed marked dependence on $\alpha$. One crucial difference
between the two systems is obvious. Condensing pseudo scalar field $\theta$ is not charged under $U(1)$-gauge group. 

The proper condensed matter interpretation of the field $\theta$ is at present not clear, but the issue  is
of great interest, especially in view of the recent measurements of the {\it superfluid} Hall effect in   
 an ultra cold gas of neutral atoms~\cite{leblanc2012}. This together with close relation between 
Hall conductivity and Hall viscosity calls for further studies and  condensed matter
interpretation. The problem is also important from cosmological point of view as the measurements 
of the Hall viscosity can possibly be used for detection of the {\it dark matter}, by observing the time variations
of $\eta_H$, i. e., in close analogy to the 
recent~\cite{kouvaris2014} proposal of the {\it dark matter} detection by the daily modulation of some parameters. 
There is however an important caveat.  Despite great number
of studies of the Hall viscosity there is no consensus how to measure this important parameter.
Time will show if the very recent proposal~\cite{abanov2014, hua15} to relate the Hall viscosity to
the  density response of the system turns out to be the feasible method to extract the former.




\acknowledgments
MR was partially supported by the grant of the National Science Center \\
$DEC-2014/15/B/ST2/00089$
and KIW by the grant DEC-2014/13/B/ST3/04451.




\begin{thebibliography}{99}

%
\def\cmp#1#2#3#4{\emph{#4}, \emph{ Commun. Math. Phys.} {\bf #1} (#3) #2}
\def\lmp#1#2#3#4{\emph{#4}, \emph{ Lett. Math. Phys.} {\bf #1} (#3) #2}
\def\hpa#1#2#3#4{\emph{#4}, \emph{ Hell. Phys. Acta} {\bf #1} (#3) #2}
\def\grg#1#2#3#4{\emph{#4}, \emph{ Gen. Rel. Grav.} {\bf #1} (#3) #2}
\def\pr#1#2#3#4{\emph{#4}, \emph{ Phys. Rev.} {\bf #1} (#3) #2}
\def\prl#1#2#3#4{\emph{#4}, \emph{ Phys. Rev. Lett.} {\bf #1} (#3) #2}
\def\prd#1#2#3#4{\emph{#4}, \emph{ Phys. Rev. D} {\bf #1} (#3) #2}
\def\prb#1#2#3#4{\emph{#4}, \emph{ Phys. Rev. B} {\bf #1} (#3) #2}
\def\pl#1#2#3#4{\emph{#4}, \emph{ Phys. Lett.} {\bf #1} (#3) #2}
\def\pla#1#2#3#4{\emph{#4}, \emph{ Phys. Lett. A} {\bf #1} (#3) #2}
\def\plb#1#2#3#4{\emph{#4}, \emph{ Phys. Lett. B} {\bf #1} (#3) #2}
\def\prep#1#2#3#4{\emph{#4}, \emph{ Phys. Reports} {\bf #1} (#3) #2}
\def\phys#1#2#3#4{\emph{#4}, \emph{ Physica} {\bf #1} (#3) #2}
\def\jcp#1#2#3#4{\emph{#4}, \emph{ J. Comput. Phys.} {\bf #1} (#3) #2}
\def\jmp#1#2#3#4{\emph{#4}, \emph{ J. Math. Phys.} {\bf #1} (#3) #2}
\def\jpm#1#2#3#4{\emph{#4}, \emph{ J. Phys. A: Math. Gen.} {\bf #1} (#3) #2}
\def\cpr#1#2#3#4{\emph{#4}, \emph{ Computer Phys. Rept.} {\bf #1} (#3) #2}
\def\cqg#1#2#3#4{\emph{#4}, \emph{ Class. Quant. Grav.} {\bf #1} (#3) #2}
\def\cma#1#2#3#4{\emph{#4}, \emph{ Computers Math. Applic.} {\bf #1} (#3) #2}
\def\mc#1#2#3#4{\emph{#4}, \emph{ Math. Compt.} {\bf #1} (#3) #2}
\def\apj#1#2#3#4{\emph{#4}, \emph{ Astrophys. J.} {\bf #1} (#3) #2}
\def\apjs#1#2#3#4{\emph{#4}, \emph{ Astrophys. J. Suppl.} {\bf #1} (#3) #2}
\def\acta#1#2#3#4{\emph{#4}, \emph{ Acta Astronomica} {\bf #1} (#3) #2}
\def\apl#1#2#3#4{\emph{#4}, \emph{ Ann. Physik. (Leipzig)} {\bf #1} (#3) #2}
\def\amjp#1#2#3#4{\emph{#4}, \emph{Am. J. Phys.} {\bf #1} (#3) #2}
\def\anp#1#2#3#4{\emph{#4}, \emph{ Ann. Phys.} {\bf #1} (#3) #2}
\def\sa#1#2#3#4{\emph{#4}, \emph{ Sov. Astro.} {\bf #1} (#3) #2}
\def\sia#1#2#3#4{\emph{#4}, \emph{ SIAM J. Sci. Statist. Comput.} {\bf #1} (#3) #2}
\def\aa#1#2#3#4{\emph{#4}, \emph{ Astron. Astrophys.} {\bf #1} (#3) #2}
\def\mnras#1#2#3#4{\emph{#4}, \emph{ Mon. Not. R. Astr. Soc.} {\bf #1} (#3) #2}
\def\npb#1#2#3#4{\emph{#4}, \emph{ Nucl. Phys. B} {\bf #1} (#3) #2}
\def\prsla#1#2#3#4{\emph{#4}, \emph{ Proc. R. Soc. London, Ser. A} {\bf #1} (#3) #2}
\def\jhep#1#2#3#4{\emph{#4}, \emph{ JHEP} {\bf #1} (#2) #3}
\def\nuc#1#2#3#4{\emph{#4}, \emph{ Nuovo Cimento B } {\bf #1} (#3) #2}
\def\ijmp#1#2#3#4{\emph{#4}, \emph{ Int. J. Mod. Phys. D} {\bf #1} (#3) #2}
\def\atmp#1#2#3#4{\emph{#4}, \emph{ Adv. Theor. Math. Phys.} {\bf #1} (#3) #2}
\def\ptps#1#2#3#4{\emph{#4}, \emph{ Prog. Theor. Phys. Suppl.} {\bf #1} (#3) #2}
\def\ptp#1#2#3#4{\emph{#4}, \emph{ Prog. Theor. Phys.} {\bf #1} (#3) #2}
\def\lmp#1#2#3#4{\emph{#4}, \emph{ Lett. Math. Phys.} {\bf #1} (#3) #2}
\def\cpam#1#2#3#4{\emph{#4}, \emph{ Comm. Pure Appl. Math.}  {\bf #1} (#3) #2}
\def\adv#1#2#3#4{\emph{#4}, \emph{ Adv. Phys.}  {\bf #1} (#3) #2}
\def\zh#1#2#3#4{\emph{#4}, \emph{ Zh. Eksp. Teor. Fiz.}  {\bf #1} (#3) #2}
\def\mplb#1#2#3#4{\emph{#4}, \emph{ Mod. Phys. Lett. B} {\bf #1} (#3) #2}
\def\jams#1#2#3#4{\emph{#4}, \emph{ J. Austral. Math. Soc. B} {\bf #1} (#3) #2}
\def\appa#1#2#3#4{\emph{#4}, \emph{ Acta Phys. Polonica A} {\bf #1}, (#3) #2}
\def\nat#1#2#3#4{\emph{#4}, \emph{Nature} {\bf #1}, (#3) #2}
\def\science#1#2#3#4{\emph{#4}, \emph{Science} {\bf #1}, (#3) #2}
\def\arcmp#1#2#3#4{\emph{#4}, \emph{Annual Rev. of Cond. Matter Physics} {\bf #1}, (#3) #2}
\def\zphys#1#2#3#4{\emph{#4}, \emph{Z. Phys.} {\bf #1}, (#3) #2}
\def\ncs#1#2#3#4{\emph{#4}, \emph{Nuovo Cimento Suppl.} {\bf #1}, (#3) #2}
\def\physb#1#2#3#4{\emph{#4}, \emph{Physica B} {\bf #1}, (#3) #2}
\def\jpcm#1#2#3#4{\emph{#4}, \emph{J. Phys.: Condens. Matter } {\bf #1}, (#3) #2}
\def\pnas#1#2#3#4{\emph{#4}, \emph{Proc. Nat. Academy Sciences} {\bf #1}, (#3) #2}
\def\sssr#1#2#3#4{\emph{#4}, \emph{Izv. Akad Nauk SSSR, ser. fiz.} {\bf #1}, (#3) #2}




\def\hepph#1#2{{ hep-ph }{#1} (#2)}
\def\hepth#1#2{{ hep-th }{#1} (#2)}
\def\grqc#1#2{{ gr-qc }{#1} (#2)}
\def\ibid#1#2#3#4{\emph{#4}, {\it ibid.} {\bf #1} (#3) #2}
\def\conphy#1#2#3#4{\emph{#4}, \emph{Contemporary Physics} {\bf #1}, (#3) #2}
%
\bibitem{mal}
J.M.Maldacena, \atmp{2}{231}{1998}{The large-N limit of superconformal field theories and supergravity}.

\bibitem{wit98}
E.Witten, \atmp{2}{253}{1998}{Anti-de-Sitter space and holography}.
\bibitem{gub98}
S.S.Gubser, I.R.Klebanov and A.M.Polyakov, \plb{428}{105}{1998}{Gauge theory correlators from noncritical string theory}.

\bibitem{kovtun2005}
P. Kovtun, D. Son and A. Starinets, \prl{94}{111601}{2005}{Viscosity in strongly interacting quantum field theories from black hole physics}.


\bibitem{rom07} 
P.Romatscke and U.Romatschke, \prl{99}{172301}{2007}{Viscosity information from relativistic nuclear collisions: how perfect is the fluid observed at RHIC}.\\
E. Shuryak, {\it Why does the quark gluon plasma at RHIC behave as a nearly ideal fluid?}, 
Prog. Part. Nucl. Phys. \textbf{53},  273 (2004).

\bibitem{mat07}
D.Mateos, \cqg{24}{S713}{2007}{String theory and quantum chronodynamics}.

\bibitem{hartnoll2007}
S. A. Hartnoll, P. K. Kovtun, M. M\"uller, S. Sachdev, \prb{76}{144502}{2007}{
Theory of the Nernst effect near quantum phase transitions in condensed matter and in dyonic black holes}.


\bibitem{cao11}
C.Cao, E.Elliot, J.Joseph, H.Wu, J.Petricka, T.Sch\"afer, and J.E.Thomas, \science{331}{58}{2011}{Universal Quantum Viscosity in a Unitary Fermi Gas}.


\bibitem{crossno2016}
J. Crossno, J. K. Shi, K. Wang, X. Liu, A. Harzheim,
A. Lucas, S. Sachdev, P. Kim, T. Taniguchi, K. Watanabe,
T. A. Ohki, K. C. Fong, \science{351}{1058}{2016}{Observation of the Dirac fluid and the breakdown of the Wiedemann-Franz law in graphene}.

\bibitem{bandurin2016}
D. A. Bandurin, I. Torre, R. Krishna Kumar, M. Ben Shalom, A. Tomadin,
A. Principi, G. H. Auton, E. Khestanova, K. S. Novoselov, I. V. Grigorieva,
L. A. Ponomarenko, A. K. Geim, M. Polini, \science{351}{1055}{2016}{Negative local resistance caused by viscous electron backflow in graphene}.

\bibitem{moll2016}
P. J. W. Moll, P. Kushwaha, N. Nandi,
B. Schmidt, A. P. Mackenzie, \science{351}{1061}{2016}{Evidence for hydrodynamic electron flow in PdCoO2}.


\bibitem{zaanen2016}
J. Zaanen, \science{351}{1026}{2016}{Electrons go with the flow in exotic material systems}.

\bibitem{madelung1926}
E. Madelung, \zphys{40}{322}{1926}{Quantentheorie in hydrodynamischer form}.

\bibitem{dej1995}
M. J. M. de Jong and L. W. Molenkamp, \prb{51}{13389}{1995}{Hydrodynamic electron flow in high-mobility wires}.

\bibitem{eaves1999}
L. Eaves, \physb{272}{130}{1999}{A hydrodynamic description of quantum Hall effect breakdown}.

\bibitem{abanov2013}
A. G. Abanov, \jpm{46}{292001}{2013}{On the effective hydrodynamics of the fractional quantum Hall effect}.

\bibitem{dagosta2006}
R. D'Agosta and M. Di Ventra, \jpcm{18}{11059}{2006}{Hydrodynamic approach to transport and turbulence in nanoscale conductors}.

\bibitem{landau1953}
L. D. Landau, \sssr{17}{51}{1953}{On multiple production of particles during collisions of fast particles}.
\bibitem{bel56}
S.Belen'kji and L.D.Landau, \ncs{3}{15}{1956}{Hydrodynamic Theory of Multiple Production of Particles}.







\bibitem{avr95}
J. E. Avron, R. Seler, and P. G. Zograf, \prl{75}{697}{1995}{Viscosity of Quantum Hall Fluids}.


\bibitem{rea11}
N. Read and E. H. Rezayi, \prb{84}{085316}{2011}{Hall viscosity, orbital spin, and geometry: Paired superfluids and quantum Hall systems}.

\bibitem{bra12}
B. Bradlyn, M. Goldstein, and N. Read, \prb{86}{24309}{2012}{Kubo formulas for viscosity: Hall viscosity, Ward identities, and the relation with conductivity}.

\bibitem{jan06a}
 R.A. Janik and R. Peschanski, \prd{73}{045013}{2006}{Asymptotic perfect fluid dynamics as a consequence of AdS/CFT}.
 \bibitem{jan06b}
 R.A. Janik and R. Peschanski, \prd{74}{046007}{2006}{Gauge/gravity duality and thermalization of a boost- invariant perfect fluid}.

\bibitem{jan07}
R.A. Janik, \prl{98}{022302}{2007}{Viscous plasma evolution from gravity using AdS/CFT}. 

\bibitem{sar06}
O.Saremi, \jhep{10}{2006}{083}{The viscosity bound conjecture and hydrodynamics of M-2-brane theory at finite chemical potential}.

\bibitem{buc05}
A. Buchel, J. T. Liu and A. O. Starinets, \npb{707}{56}{2006}{Coupling constant dependence 
of the shear viscosity in N=4 supersymmetric Yang-Mills theory}.

\bibitem{ben06}
P. Benincasa and A. Buchel, \jhep{01}{2006}{103}{Transport properties of N=4 supersymmetric Yang-Mills 
theory at finite coupling}.
\bibitem{buc08}
A. Buchel, \npb{802}{281}{2008}{Shear viscosity of boost invariant plasma at finite coupling}.
\bibitem{buc08a}
A. Buchel, R. C. Myers, M. F. Paulos and A. Sinha, \plb{669}{364}{2008}{Universal holographic 
hydrodynamics at finite coupling}.
\bibitem{buc08b}
A.Buchel, \plb{665}{298}{2008}{Shear viscosity of CFT plasma at finite coupling}.
\bibitem{buc08c}
A.Buchel, \npb{803}{166}{2008}{Resolving disagreeement for $\eta/s$ in a CFT plasma at finite coupling}.
\bibitem{mye09}
R.C.Myers, M.F.Paulos and A.Sinha, \prd{79}{041901}{2009}{Quantum corrections to $\eta/s$}.
\bibitem{cre11}
S.Cremonini, \mplb{25}{1867}{2011}{The shear viscosity to entropy ratio: a status report}.

\bibitem{bri08}
M.Brigante, H.Liu, R.C.Myers, S.Shenker and S.Yaida, \prd{77}{126006}{2008}{Viscosity bound violation in higher derivative gravity}.
\bibitem{bri08a}
M.Brigante, H.Liu, R.C.Myers, S.Shenker and S.Yaida, \prl{100}{191601}{2008}{The viscosity bound violation and causality violation}.


\bibitem{reb12}
A.Rebhan and D.Steineder, \prl{108}{021601}{2012}{Violation of the holographic bound in a strongly coupled anisotropic plasma}.
\bibitem{jai15}
S.Jain, N.Kundu, K.Sen, A.Sinha and S.P.Trivedi, \jhep{01}{2015}{005}{A strongly coupled anisotropic fluid from dilaton driven holography}.
\bibitem{ge15}
X.H.Ge, Y.Ling, C.Niu and S.J.Sin, \prd{92}{106005}{2015}{Thermoelectric conductivities, shear viscosity and stability in an anisotropic linear axion model}.
\bibitem{alb16}
L.Alberte, M,Baggioli and O.Pujolas, \hepth{1601.03384}{2016}{Viscosity bound violation in holographic solids and the viscoelastic response}.
\bibitem{sad16}
M.Sadeghi and S.Parvizi, \cqg{33}{035005}{2016}{Hydrodynamics of a black brane in Gauss-Bonnet massive gravity}.






\bibitem{sar12}
O. Saremi and D.T. Son, \jhep{04}{2012}{011}{Hall viscosity from gauge/gravity duality}.
\bibitem{che12}
J.W.Chen, N.E.Lee,  D.Maity and W.Y.Wen, \plb{713}{47}{2012}{A holographic model for Hall viscosity}.
\bibitem{roy14}
D.Roychowdhury, \jhep{10}{2014}{015}{Hall viscosity to entropy ratio in higher derivative theories}.

\bibitem{cai12}
R.G.Cai, T.J.Li, Y.H.Qi and Y.L.Zhang, \prd{86}{086008}{2012}{Incompressible Navier-Stokes equations from Einstein gravity with Chern-Simons term}.
\bibitem{zou14}
D.C.Zou and B.Wang, \prd{89}{064036}{2014}{Holographic parity violating charged fluid dual to Chern-Simons modified gravity}.

\bibitem{liu13}
H.Liu, H.Ooguri and B.Stoica, \prl{110}{211601}{2013}{Spontaneous generation of angular momentum in holographic theories}.
\bibitem{liu14}
H.Liu, H.Ooguri and B.Stoica, \prd{89}{106007}{2014}{Angular momentum generation by parity violation}.
\bibitem{liu14a}
H.Liu, H.Ooguri and B.Stoica, \prd{90}{086007}{2014}{Hall viscosity and angular momentum in gapless holographic models}.
\bibitem{son14}
D.T.Son and C.Wu, \jhep{07}{2014}{076}{Holographic spontaneous parity breaking and the emergent Hall viscosity and angular momentum}.

\bibitem{erd13}
J. Erdmenger, D. Fernandez, and H. Zeller, \jhep{04}{2013}{049}{New transport properties of anisotropic holographic superfluids}.
\bibitem{jen12}
K.Jensen, M.Kaminski, P.Kovtun, R.Meyer, A.Ritz, and A.Yarom, \prl{109}{101601}{2012}{Towards hydrodynamics without an entropy current}.
\bibitem{jen12a}
K.Jensen, M.Kaminski, P.Kovtun, R.Meyer, A.Ritz, and A.Yarom, \jhep{05}{2012}{102}{Parity-violating hydrodynamics in 2+1 dimensions}.
\bibitem{hua15}
B. Huang, \prb{91}{235101}{2015}{Hall viscosity revealed via density response}.





\bibitem{nak14}
{\L}.Nakonieczny and M.Rogatko, \prd{90}{106004}{2014}{Analytic study on backreacting holographic superconductors with dark matter sector}.
\bibitem{nak15}
{\L}.Nakonieczny, M.Rogatko and K.I. Wysoki\'nski, \prd{91}{046007}{2015}{Magnetic field in holographic superconductors with dark matter sector}.
\bibitem{nak15a1} \L{}. Nakonieczny, M. Rogatko and K.I. Wysoki\'nski, \prd{92}{066008}{2015}{Analytic investigation of holographic phase transitions influenced by
dark matter sector}.
\bibitem{rog15} M. Rogatko, K.I. Wysoki\'nski, {\it P-wave holographic superconductor/insulator phase
transitions affected by dark matter sector}, \hepth{1508.02869}{2015}.
\bibitem{rog15a} 
M. Rogatko, K.I. Wysoki\'nski, \jhep{12}{2015}{041}{Holographic vortices in the presence of dark matter sector}.

\bibitem{pen15a}
Y.Peng, \plb{750}{420}{2015}{Holographic entanglement entropy in superconductor phase transition with dark matter sector}.
\bibitem{pen15b}
Y.Peng, Q.Pan and Y.Liu, {\it Holographic insulator/superconductor phase transition model with dark matter sector away from the probe limit}, \hepth{1512.08950}{2015}.






\bibitem{reg15}
M.Regis, J.Q.Xia, A.Cuoso, E.Branchini, N.Fornengo, and M.Viel, \prl{114}{241301}{2015}{Particle Dark Matter Searches Outside the Local Group}.
\bibitem{ali15}
Y.Ali-Haimoud, J.Chluba, and M.Kamionkowski, \prl{115}{071304}{2015}{Constraints on Dark Matter Interactions with Standard Model Particles from Cosmic Microwave Background Spectral Distortions}.

\bibitem{foo15}
R.Foot and S.Vagnozzi, \prd{91}{023512}{2015}{Dissipative hidden sector dark matter}.
\bibitem{foo15a}
R.Foot and S.Vagnozzi, \plb{748}{61}{2015}{Diurnal modulation signal from dissipative hidden sector dark matter}.


\bibitem{bra14}
J.Bramante and T.Linden, \prl{113}{191301}{2014}{Detecting dark matter with imploding pulsars in the galactic center}.
\bibitem{ful15}
J.Fuller and C.D.Ott, \mnras{450}{L71}{2015}{Dark-matter-induced collapse of neutron stars: a possible link between fast radio bursts and missing pulsar problem}.
\bibitem{lop14}
I.Lopes and J.Silk, \apj{786}{25}{2014}{A particle dark matter footprint on the first generation of stars}.
\bibitem{nak12}
A.Nakonieczna, M.Rogatko, and R.Moderski, \prd{86}{044043}{2012}{Dynamical collapse of charged scalar field in phantom gravity}.
\bibitem{nak15a}
A.Nakonieczna, M.Rogatko, and L.Nakonieczny, \jhep{11}{2015}{016}{\it Dark matter impact on gravitational collapse of an electrically charged scalar field}.


\bibitem{ger15}
A.Geringer-Sameth and M.G.Walker, \prl{115}{081101}{2015}{Indication of Gamma-Ray Emission from the Newly Discovered Dwarf Galaxy Reticulum II}.
\bibitem{til15}
K.Van Tilburg, N.Leefer, L.Bougas, and D.Budker, \prl{115}{011802}{2015}{Search for Ultralight Scalar Dark Matter with Atomic Spectroscopy}.

\bibitem{integral}
P.Jean {\it et al.}, \aa{407}{L55}{2003}{Early SPI/INTEGRAL measurements of 511 keV line emission from the 4th quadrant of the Galaxy}.
\bibitem{atic}
J.Chang {\it et al.}, \nat{456}{362}{2008}{An excess of cosmic ray electrons at energies of 300-800 GeV}.
\bibitem{pamela}
O.Adriani {\it et al.} (PAMELA Collaboration), \nat{458}{607}{2009}{An anomalous positron abundance in cosmic rays with energies 1.5-100 Gev}.
\bibitem{muon}
G.W.Bennett {\it et al.}, \prd{73}{072003}{2006}{Final report of the E821 muon anomalous magnetic moment measurement at BNL}.

\bibitem{massey15a}
D.Harvey, R.Massey, T.Kitching, A.Taylor and E.Tittley, \science{347}{1462}{2015}{The nongravitational interactions of dark matter in colliding galaxy clusters}.
\bibitem{massey15b}
R.Massey {\it et al.}, \mnras{449}{3393}{2015}{The behaviour of dark matter associated with four bright cluster galaxies in the 10 kpc core of Abell 3827}.
\bibitem{babar14}
J.P.Lees et al., \prl{113}{201801}{2014}{Search for a Dark Photon in $e^+ e^-$ Collisions at BABAR}.




\bibitem{jac03}
R.Jackiw and S.Y.Pi, \prd{68}{104012}{2003}{Chern-Simons modification of general relativity}.
\bibitem{ale09}
S.Alexander and N.Yunes, \prep{480}{1}{2009}{Chern-Simons modified general relativity }.
\bibitem{rog13}
M.Rogatko, \prd{88}{024051}{2013}{Uniqueness of charged static asymptotically flat black holes in dynamical Chern-Simons gravity}.
\bibitem{har08}
S.A.Hartnoll, C.P.Herzog, and G.T.Horowitz, \jhep{12}{2008}{015}{Holographic superconductors}.
\bibitem{fau09}
T.Faulkner, H.Liu, J.McGreevy, and D.Vegh, \prd{83}{125002}{2011}{Emergent quantum criticality, Fermi surfaces, and AdS2}.

\bibitem{bri11} 
Y.Brihaye and B.Hartmann, \prd{80}{123502}{2009}{Effect of dark strings on semilocal strings}.
\bibitem{kra06}
P.Kraus and F.Larsen, \jhep{01}{2006}{022}{Holographic gravitational anomalies}.
\bibitem{gru08}
D.Grumiller, R.B.Mann, and R.McNees, \prd{78}{081502}{2008}{Dirichlet boundary value problem for Chern-Simons modified gravity}.
\bibitem{fef85}
C.Fefferman and C.R.Graham, {\it Conformal invariants}, in {\it Ellie Cartan et les Math\'ematiques d'aujourd'hui} (Ast\'erisque, 1985)~ 95.


\bibitem{bha08}
S.Bhattacharya, V.E.Hubeny, S.Minwalla and M.Rangamani, \jhep{02}{2008}{045}{Nonlinear fluid dynamics from gravity}.




\bibitem{raa08}
M. Van Raamsdonk, \jhep{05}{2008}{106}{Black hole dynamics from atmospheric science?}.

\bibitem{kle99}
I.R.Klebanov and E.Witten, \npb{556}{89}{1999}{AdS/CFT correspondence and symmetry breaking}.


\bibitem{her06}
T.Hertog, \prd{74}{045002}{2010}{Towards a novel no-hair theorem for black holes}.


\bibitem{refe} {\it We are gratefull to the Referee for asking the question, which contributed
to the clarification of this important point.}



\bibitem{iqb10}
N.Iqbal, H.Liu, M.Mezei, and Q.Si, \prd{82}{045002}{2010}{Quantum phase transitions in holographic models of magnetism and superconductors}.

\bibitem{har01}
S.de Haro, K.Skendris and S.N.Solodukhin, \cmp{217}{595}{2001}{Holographic Reconstruction of Spacetime and Renormalization in the AdS/CFT correspondence}.



\bibitem{leblanc2012}
L. J. LeBlanc, K. Jimenez-Garcia, R. A. Williams, M. C. Beeler, A. R. Perry, W. D. Phillips, and I. B. Spielman,  \pnas{109}{10811}{2012}{Observation of a superfluid Hall effect}.

\bibitem{kouvaris2014}
C. Kouvaris and I. M. Shoemaker, \prd{90}{095}{2014}{Daily modulation as smoking gun
of dark matter with significant stopping rate}.



\bibitem{abanov2014}
A. G. Abanov and A. Gromov, \prd{90}{014435}{2014}{Electromagnetic and gravitational responses of two-dimensional noninteracting electrons in a background magnetic field}.








\end{thebibliography}
\end{document}